\documentclass[10pt,letterpaper]{article}
\usepackage{opex3}
\usepackage{cite}

\begin{document}

\title{Tuning resonant interaction of
orthogonally polarized solitons and
dispersive waves with the soliton power}

\author{A.V. Yulin$^1$, L.R. Gorj\~{a}o$^1$, R. Driben$^{2}$, D.V. Skryabin$^{1,3}$}

\address{$^1$1Centro de F\'{i}sica Te\'{o}rica e Computacional,
Universidade
de Lisboa, Ave. Prof. Gama Pinto 2, Lisboa 1649-003, Portugal\\
$^2$Department of Physics
and CeOPP, University of Paderborn, Warburger Str. 100, D-33098
Paderborn, Germany\\
$^3$Department of Physics, University of Bath, BA2 7AY, United Kingdom}

\email{  } 



\begin{abstract}
We demonstrate that the relatively small power induced changes
in the soliton wavenumber comparable with splitting of the effective indexes
of the orthogonally polarized waveguide modes result in significant changes
of the efficiency of the interaction between solitons and dispersive waves
and can be used to control energy transfer between the soliton and newly
generated waves and to delay or accelerate solitons.
\end{abstract}







Interaction of optical solitons with dispersive waves (DWs) is
an active research topic, which besides its fundamental significance  plays an important role in understanding of supercontinuum (SC) generation \cite{skr}.
In particular, it has been demonstrated that at advanced stages of SC generation
interaction of solitons with strong DWs can lead to  significant changes of the soliton and DW frequencies and energies \cite{Nature_Gorbach,JOSAB_Judge,OE2010_Driben,Sci_Demircan,OL_Driben,OL2_Driben,PRE_Driben}.
A  related effect is the long range soliton-soliton interaction mediated by  DWs
trapped between the solitons and leading to either acceleration or deceleration of solitons
and significant spectral reshaping  happening in the course of  propagation \cite{OE_yulin,OE_Driben}.
Interaction of solitons with a strong DW has been also proposed  for use in  all optical switching \cite{Steinmayer} and
for SC generation \cite{add1,add2}.

A fundamental  process behind the effects observed in the interaction of solitons with DWs
is the four-wave mixing (FWM) of DWs with the selected frequency components of
the soliton spectrum \cite{skr}. This type of FWM is mediated either by the nonlinear
refractive index change induced by the soliton intensity (cross-phase modulation (XPM) effect)
or by the FWM term  (square of the soliton field times the complex conjugated DW field) \cite{PRL_Efimov,OL_Efimov,skr}.
The DW-soliton interaction through the XPM and FWM processes can be
referred to as the phase insensitive and phase sensitive processes, respectively.
Efficiency of the frequency conversion happening in the soliton-DW interaction depends
on many factors and potential avenues of its enhancement, possible novel effects and practical
applications have not been explored  in sufficient details.

Efficiency of the soliton-DW interaction can be trivially boosted through the  increase of
the amplitudes of the involved waves, so that the nonlinear mixing terms become larger.
Less obvious, but more sensitive mechanism is taking an advantage of the power dependence
of the phase matching (PM) conditions, so that, e.g., the signal waves generated through the DW-soliton interaction
can either appear or disappear together with PM itself.
The nonlinear shift of the soliton wavenumber is not a significant factor in the
PM condition for the soliton to emit a dispersive wave (Cherenkov radiation)
and can be disregarded \cite{skr}. PM condition for the phase insensitive (XPM induced)
scattering of an externally applied DW on a soliton
simply does not depend on the nonlinear shift of the soliton phase
\cite{OL_yulin,PRE_Skryabin,Nature_Gorbach}.
However, the situation is different if one considers the phase
sensitive interaction \cite{OL_Efimov,skr}. In this case the nonlinear phase shifts can become
comparable with the linear wavenumber splitting of, e.g., orthogonally polarized modes,
and hence the soliton-DW scattering processes with the PM conditions
critically depending on the soliton power become feasible.
While the power dependance has been previously reported \cite{OL_Efimov,skr}, its impact on the dispersive wave scattering and amplification has not been addressed  so far.

This work aims to target several problems related to the above conjecture.
In particular,  we will demonstrate that  the presence or absence of the frequency
conversion achieved in the phase sensitive interaction of the orthogonally polarized solitons and DWs
critically depend on the soliton intensity and can be either enhanced
or completely suppressed through relatively small changes in the
soliton power. The signs of the soliton acceleration  and  of the
frequency drift induced by DWs also can be controlled through the soliton power.

Propagation of polarized light in optical waveguides can be described
using the following dimensionless equations \cite{PRL_Lu}:
\begin{equation}
\left[i\partial_z + \hat D_{x, y}\right] A_{x,y} +
\left[|A_{x,y}|^2 +\frac{2}{3}|A_{y,x}|^2\right]A_{x,y} +  \frac{1}{3}A_{y,x}^2A_{x,y}^{*} = 0,
\label{eq1}
\end{equation}
where $A_{x,y}$ are the  amplitudes of the orthogonally polarized  modes.
$\hat D_{x}(i\partial_t)=\frac{1}{2}\partial^2_{t}$ and
$\hat D_{y}(i\partial_t)=\beta_0 + i\beta_1\partial_t + \frac{\beta_2}{2}\partial^2_{t}$
are the dispersion operators. $\beta_0=(\beta^{(0)}_y-\beta^{(0)}_x)L_d$
is the normalised difference of the propagation constants of the two modes,
$\beta_1=(\beta^{(1)}_y-\beta^{(1)}_x)L_d$ is the difference of their inverse group velocities,
and $\beta_2=\beta^{(2)}_y/\beta^{(2)}_x$ is the ratio of their group velocity dispersion (GVD) coefficients.
We will assume hereafter that $\beta_2<0$, so that  GVD  is anomalous in the x-component and
normal in the y-component. Thus $x$- and $y$-components
are natural hosts for the soliton and dispersive wave pulses, respectively.
Time $t$ is normalized to the pump pulse duration,
propagation distance $z$ is measured in units of the GVD length, $L_d$, of the $x$-mode,
$|A_{x,y}|^2$ are normalized to the soliton power, making the nonlinear length  equal to the GVD length.
We have not included Raman nonlinearity, assuming that our results can  be applied for the sufficiently long pulses
in optical fibers and in the context of semiconductor waveguides \cite{silicon,algas}, where
the Raman effect on even ultrashort  pulses is negligible.

Input ($z=0$) conditions used through most of our work
is the soliton in the x-polarization and DW in the y-polarization:
$A_x=\sqrt{2q}~sech\left( \sqrt{2q }t \right)$,
$A_y=B ~sech\left( (t-t_0)/T \right)e^{i \omega_i t}$,
where $q$ is the soliton parameter proportional to its peak power and giving the nonlinear shift of the
soliton wavenumber $e^{iqz}$, $B$ is the amplitude and  $T$ is the duration of the DW pulse,
$t_0$ is the delay and $\omega_i$ is the frequency offset between the interacting pulses.
Through out  this work we consider $\beta_0>0$, which corresponds to the phase velocity in the soliton component being greater than the one in the dispersive wave component, so that soliton is the 'fast wave'.
Note that we restrict our considerations to the parameter values where
the fast wave soliton  is stable with respect to the exponential growth of the orthogonally polarised
small amplitude perturbations; see Fig. 2(c), Ref. \cite{fang} and references therein.
PM condition governing DW-soliton mixing are \cite{OL_yulin,OL_Efimov,PRE_Skryabin}
\begin{equation}
\beta(\omega)=\beta(\omega_i) \label{pm1}
\end{equation}
for the phase insensitive process and
\begin{equation}
\beta(\omega)=2q-\beta(\omega_i) \label{pm2}
\end{equation}
for the phase sensitive one, where $\beta(\omega)=\hat D_y(\omega)=\beta_0+\beta_1\omega-{1\over 2}\beta_2\omega^2$.
Roots of Eqs. (\ref{pm1}),(\ref{pm2}) give frequencies of the DWs transmitted through, $\omega_t$,
and reflected by, $\omega_r$, a soliton. Note, that  the multiple transmitted
and reflected waves are allowed to coexist.

\begin{figure}[h]
\centering\includegraphics[scale=.5]{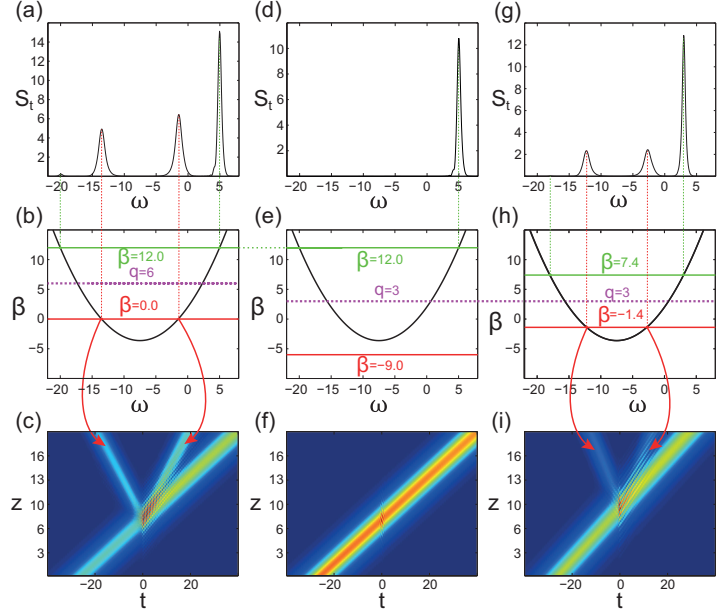}
\caption{(a)-(c) show the output spectrum (a), graphical representation of the phase matching conditions (b) and
 evolution of the y-component of the field (c). The incident DW has $\omega_i=5$ and $B=0.0333$,
the soliton parameter $q=6$. Crossings of the parabola in (b) with the green and red horizontal lines correspond to the phase insensitive and phase sensitive resonances, respectively.
(d)-(f) and (g)-(i) are the same as (a)-(c), but for  $\omega_i=5$, $q=3$, $B=0.0333$ and
$\omega_i=3$,  $q=3$, $B=0.0333$, respectively. Dotted horizontal lines indicate the soliton wavenumber.
Here and in all the other figures: $\beta_0=2$, $\beta_1=1.5$ and $\beta_2=-0.2$.}
\label{fig1}
\end{figure}

\begin{figure}[h]
\centering\includegraphics[scale=0.6]{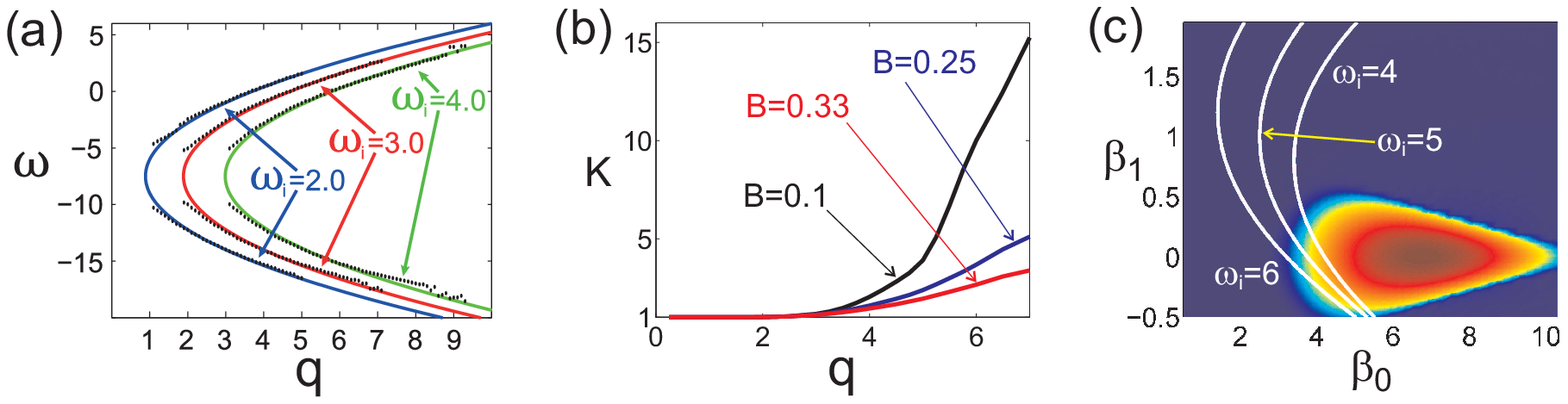}
\caption{(a) Resonance frequencies for the phase sensitive process as a function of
the soliton parameter $q$. Analytical solutions are given by the solid lines,
the black dots show the frequencies  obtained from numerical simulations.
(b)   The amplification coefficient $K$ defined as a ratio of the photon numbers in the transmitted waves at frequency $\omega_i$ to the photon number in the incident wave; $\omega_i=3$. (c) Domains  on the left from the parabolas give the values of  $\beta_0$ and $\beta_1$ where the phase sensitive conditions can be satisfied for the suitable  resonance frequencies (see Fig. 2(a)). The red-orange color indicates the region of the polarization instability of the soliton solutions for $q=5$.}
\label{fig2}
\end{figure}

Figs. 1(a-c) illustrate a typical scattering event.
An incident DW  with $\omega_i=5$ propagates to the right showing some dispersion.
When it overlaps with the soliton ($q=6$) it gets  reflected into a wave with
$\omega_r=-14$ and splits on transmission into two waves with
$\omega_t=\omega_i$ and $\omega_t=-0.9$, see Figs. 1(a,c).
Fig. 1(b) shows the associated phase matching diagram.
The phase insensitive resonances, Eq. (\ref{pm1}), are given by
the green line. These do not depend on $q$ and hence on the soliton power.
One of these resonances corresponds to the incident wave
and the other one is negligibly weak in this instance.
The phase sensitive resonances, Eq. (\ref{pm2}), are given by the red line
and they correspond to the one reflected and one transmitted waves. Reducing the soliton power,
i.e. reducing $q$, the phase sensitive resonances tend towards degeneracy at the bottom of the
parabolic dispersion of the linear waves, simultaneously, the power of the scattered waves drops down.
The point of the exact degeneracy corresponds to the propagation of the transmitted and reflected waves
parallel to the soliton interface.
The phase sensitive processes are phase matched for
$q>q_{cr}$, where $q_{cr}=\beta_0+ \left( \beta_1 \omega_i -\frac{\beta_2 \omega_i^2}{2} + \frac{\beta_1^2}{2\beta_2} \right)/2$.
The cases of $q<q_{cr}$  and $q$ greater than, but close to, $q_{cr}$
are illustrated in Figs. 1 (d-f) and (g-i), respectively.
Phase sensitive resonances as functions of $q$ for several values of $\omega_i$ are shown in Fig. 2(a), where
one can see that $q_{cr}$ increases with $\omega_i$.
Thus, tuning the soliton power one can not only shift the resonant frequencies
of the phase sensitive FWM-mediated scattering,
but also suppress the resonance completely.
Boundaries separating the areas in the $(\beta_0,\beta_1)$-plane where phase matching is possible from the ones where it is not are shown in Fig. 2(c),
together with the threshold of the polarization instability of the soliton. One can see that there is a limit on how large $\beta_0$
is allowed to be for the phase matching to be realizable.

\begin{figure}[h]
\centering\includegraphics[scale=.5]{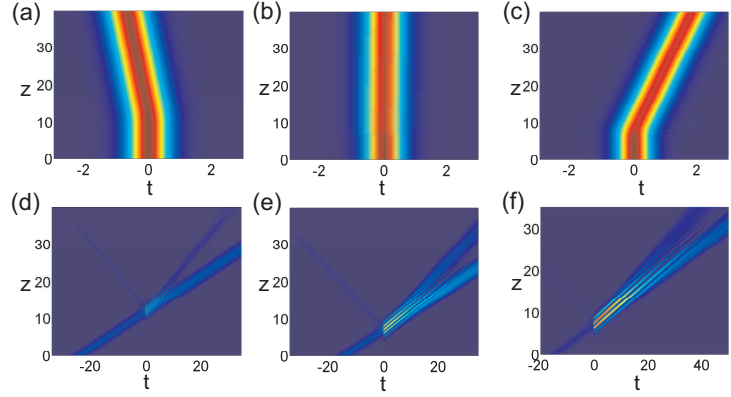}
\caption{Evolution of the total intensity $|A_x|^2$ (a-c) and $|A_y|^2$ (d-f) resulting from the soliton
collision with a  DW: (a,d) $B=0.14$, $\omega_i=3$, $q=3.5$;  (b,e) $B=0.14$,  $\omega_i=3$, $q=4.35$;
 (c,f) $B=0.14$,  $\omega_i=3$, $q=5$.}
\label{fig3}
\end{figure}

\begin{figure}[h]
\centering\includegraphics[scale=.4]{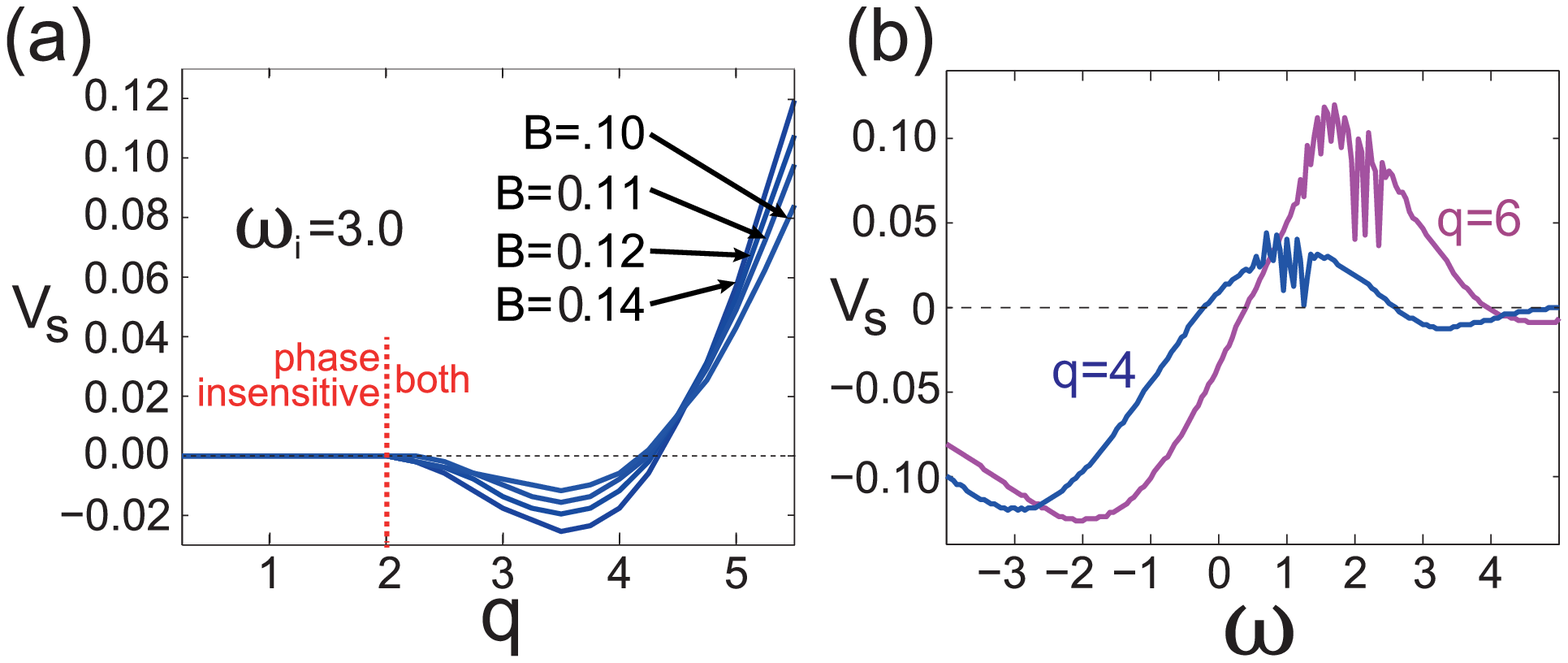}
\caption{(a) Changes in the soliton group velocity
after interaction with the DW having $\omega_i=3$
as a function of the soliton parameter $q$.
The red dashed line indicates $q_{cr}$ when the phase sensitive resonances
appear. (b) Changes in the soliton group velocity
after interaction with the DW as a function of the DW frequency $\omega_i$.
The paramerers of the solitons are $q=4$ and $q=6$, the amplitude
of the incident wave is $B=0.5$.}
\label{fig4}
\end{figure}

One should also note significant amplification of the wave transmitted without any change in frequency, see Fig. 1(c).
While theoretical understanding of this potentially useful effect has not been developed yet, we can
quantify it numerically calculating relative change in the photon number (amplification coefficient) $K$
of the pulse launched at the $\omega_i$ frequency:
$K=E(z\rightarrow \infty)/E(z=0)$, where $E(t)=\int |A_y|^2dt$.
Plots of $K$ vs the soliton parameter $q$ are shown in Fig.~\ref{fig2}(b) revealing an order
of magnitude energy change, thereby suggesting to look for the possible
role of this effect in shaping polarization dependent supercontinuum spectra \cite{PRL_Lu}. In this panel the amplification coefficient is
defined as a ratio of the total photon number in both transmitted waves to the photon number in the incident wave is shown by dashed curves.

Alongside with the generation of DWs with new frequencies,
the DW-soliton interaction leads to the appreciable impact of DWs on the soliton itself,
resulting in the change of the soliton momentum associated with
the  shift of the soliton frequency and bending of its spatio-temporal trajectory  \cite{PRE_Skryabin,Steinmayer}.
If the  phase insensitive scattering associated with the DW reflection dominates over the other scattering channels,
then the associated changes of the soliton frequency and velocity are such that
the soliton trajectory bends towards the incident DW \cite{OE_yulin}. For example, if DW hits
a soliton from the right then  the soliton trajectory bends to the left.
If  one takes two solitons and arranges for the reflected DW to bounce between them, the net force exerted
on the solitons results in attraction \cite{OE2010_Driben,Sci_Demircan,OE_yulin,OE_Driben,Steinmayer}.
However, when the phase sensitive scattering processes dominates, which happens for some $q$ above $q=q_{cr}$,
we  observe that for the sufficiently large soliton amplitudes the soliton trajectory
bends away from the incident wave, see Fig.~\ref{fig3}(c,f).
If the phase sensitive resonances are eliminated ($q<q_{cr}$),
then the soliton trajectory bends towards the incident wave as in
the scalar case, see Figs. \ref{fig3}(a,d).  Naturally, there exist parameters,
when the phase sensitive and phase insensitive processes balance each other
so that the soliton frequency does not change, see Figs. 3(b,e). Dependencies of the soliton velocity shift induced by the scattering of DW
vs the soliton parameter $q$ and vs frequency, $\omega_i$,
of the incident wave containing intervals of positive and negative velocities are shown in Fig. 4.

To conclude we briefly summarize the main results reported in the paper. It is shown that in vector case the frequencies of the radiation generated by four wave mixing of DW with solitons can be very sensitive not only to the frequency but also to the intensity of the solitons. It was demonstrated that the phase sensitive resonant scattering can be completely suppressed in case of solitons of low intensity. Studying the recoil from the scattering of the DW on solitons we demonstrated that the soliton trajectory can bend either to the left or to the right depending on the intensity of the soliton. In other words the resonant scattering changes the frequency of solitons and the sign of the frequency shift can depend on the soliton intensity. Another interesting effect considered in the paper is the amplification of DW happening due to four-wave mixing between the solitons and the DW. The reported effects open new possibilities to control optical soliton by the dispersive waves and can potentially be important for practical applications.

\section*{Acknowledgements}
We acknowledge support by the  FCT (Portugal) grant PTDC/FIS/112624/2009
and by the UK EPSRC grant EP/G044163/1.

\end{document}